\documentclass[aps,prbbib,twocolumn,epsf,showpacs]{revtex4}
\usepackage{graphicx}
\begin{document}

\title{How edge states are destroyed in disordered mesoscopic samples?}
\author{Zhenhua Qiao$^{1}$, Jian Wang$^{1,*}$, Qingfeng Sun$^{2}$, Hong Guo$^3$}
\address{
1. Department of Physics and the center of theoretical and
computational physics,\\
The University of Hong Kong, Pokfulam Road, Hong Kong, China\\
2. Institute of Physics, Chinese Academy of Sciences, Beijing, P.R.
China\\
3. Department of Physics, McGill University, Montreal, PQ, H3A 2T8,
Canada }
\begin{abstract}
We report theoretical investigations on how edge states are
destroyed in disordered mesoscopic samples by calculating a ``phase
diagram" in terms of energy versus disorder strength $(E,W)$, and
magnetic field versus disorder strength $(B,W)$, in the integer
quantum Hall regime. It is found that as the disorder strength $W$
increases, edge states are destroyed one by one if transmission
eigen-channels are used to characterize the edge states. Near the
insulating regime, transmission eigen-channels are closed one by one
in the same order as edges states are destroyed. To identify those
edge states which have survived disorder, we introduce a generalized
current density that can be calculated and visualized.
\end{abstract}


\maketitle

When a two-dimensional \emph{mesoscopic} sample is subjected to
external magnetic field, peculiar electronic states --- edge states,
may be established at the boundaries of the
sample\cite{Helperin82,buttiker88}. Classically, Lorentz force
pushes electrons toward the sample boundary and electron
trajectories become skipping orbits. Edge states can be considered
as the quantum version of skipping orbits\cite{buttiker88}.
Importantly, edge states in mesoscopic samples provide necessary
density of states (DOS) between the Landau levels, integer quantum
Hall effect (IQHE) can therefore occur in the clean sample
limit\cite{buttiker88}. In contrast, for infinitely large samples, a
degree of disorder in the sample appears necessary which provides
DOS in between Landau levels to stablize the Fermi energy for
IQHE\cite{prange-book}. Nevertheless, increasing disorder will
eventually destroy IQHE and how does this happen has been an
important issue attracting numerous studies.

\begin{figure}
\includegraphics[width=9cm,height=6.5cm]{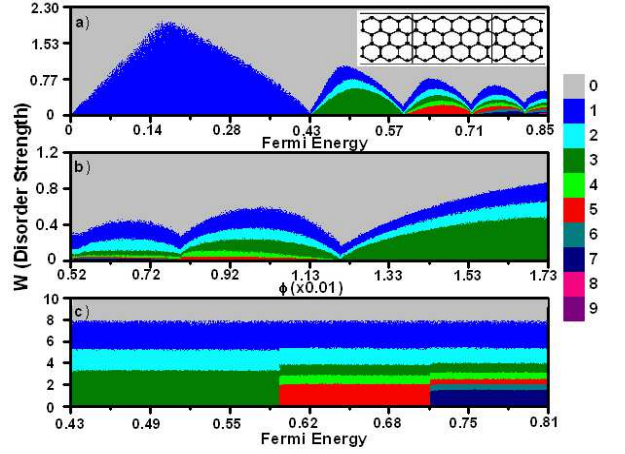}
\caption{(color online) (a,b) are ``Phase diagram" of edge states of
graphene for energies near the Fermi level (Dirac electrons): (a) in
the $(E,W)$ plane; (b) in the $(B,W)$ plane. (c) The order of
closing of eigen-channels at large disorder. Color coding are for
number of eigen-channels.} \label{fig1}
\end{figure}

Here we address the disorder issue for {\it mesoscopic} samples,
namely how edge states are destroyed by disorder in the IQHE regime
and, at a fixed filling factor $\nu > 1$, are edge states destroyed
all at once or one by one. These important questions provide insight
to the IQHE phase diagram for mesoscopic samples, and may shed light
to similar problems in samples of infinite size. We address these
questions by extensive calculations on a mesoscopic graphene system
and a square lattice model (see inset of Fig.\ref{fig1}a) to map out
a ``phase diagram" of edge states in the presence of disorder. Here
the ``phases" in the ``phase diagram" denote quantum states and no
phase transitions are implied between these states. We use
transmission eigen-channels to characterize edge states (see below),
and we found that they are destroyed one by one. At large enough
disorder, the system becomes an insulator and transmission
eigen-channels are closed one by one in the same order as the edge
states are destroyed.

We begin by discussing our definition of edge states as well as the
way to visualize them. In transport theory, for each incoming
channel of a semi-infinite lead $\alpha$ whose wave function is
$|W_{\alpha,m}>$ where $m=1,2,\cdots N$ denotes one of the $N$
channels, one solves a scattering problem. $|W_{\alpha,m}>$ is an
eigen-channel of lead-$\alpha$ but not the entire device. To find
the eigen-channels of the entire device, we
diagonalize\cite{buttiker1} the transmission matrix ${\cal T}$ by a
unitary transformation $U$, for a two probe device having scattering
matrix $S_{RL}$, ${\cal T}=S^\dagger_{RL}S_{RL}$. Mathematically,
this means acting $U$ on the incoming channels $|W_{\alpha,m}>$
(which is a column vector with N components) to obtain a new set of
orthogonal incoming modes $|{\bar W}_{\alpha m}>=\sum_{n}
|W_{\alpha,n}>U_{nm}$. Once done, $|{\bar W}_{\alpha,m}>$ is an
\emph{eigen-state} of the entire device (leads plus scattering
region). In other words, if an incoming electron comes at state
$|{\bar W}_{\alpha,m}>$, it will traverse the entire device without
mixing with any other eigen-state $|{\bar W}_{\alpha,m'}>$. This
way, the resulting transmission matrix ${\bar {\cal T}}=U^\dagger
{\cal T} U$ is diagonal. In the presence of a strong magnetic field
$B$, it is therefore natural to identify $|{\bar W}_{\alpha m} (B)>$
as edge states because they are the eigen-states or eigen-channels
of the entire device sample. How to visualize edge states in the
IQHE regime? This may be achieved by plotting the current density in
real space. The subtle issues of current density in IQHE has been
discussed in Ref.\onlinecite{Hirai}. In our work where there are
disorder in the sample, the eigen-channels provide a convenient way
to define a generalized current density for each channel --- since
the eigen-channels do not mix. Clearly, the total transport current
is obtained by integrating current density along any cross-section
perpendicular to the transport direction.

The total transmission coefficient is obtained from the $N\times N$
Hermitian transmission matrix ${\cal T}=\{t_{ij}\}$. Applying an
unitary transformation $U$\cite{buttiker1}, we obtain ${\bar {\cal
T}}=U^\dagger {\cal T} U=\{t_i \delta_{ij}\}$, which is diagonal
with elements $t_i$. The transmission coefficient $t_i$ of
eigen-channel $i$ is a linear combination of $t_{ij}$. Using
conventional current density ${\bf J}_{ii}=(ie\hbar/2m) (\psi_i
\nabla \psi_i^* - \psi^*_i \nabla \psi_i)-(e^2/m){\bf A} |\psi_i|^2$
with $i=1,2,\cdots N$, it is easy to show $(e^2/h) ~ t_{ii} = \int
{\bf J}_{ii} \cdot d{\bf s}$ in the linear bias regime, where
$t_{ii}$ is the diagonal element of matrix ${\cal T}$. In order to
find an \emph{eigen-current density} ${\bf J}_{i}$ such that
\begin{equation}
\frac{e^2}{h}~ t_{i} = \int {\bf J}_{i} \cdot d{\bf s}\ ,
\label{eq1}
\end{equation}
we define a generalized complex current density ${\bf J}_{ij}$ so
that $(e^2/h) ~ t_{ij} = \int {\bf J}_{ij} \cdot d{\bf s}$. The
unitary transformation on the incoming wave function
$|W_{\alpha,m}>$ suggests the following definition:
\begin{equation}
{\bf J}_{ij} = \frac{ie\hbar}{2m} (\psi_j \nabla \psi_i^* - \psi^*_i
\nabla \psi_j) -\frac{e^2}{m}{\bf A} \psi_j \psi^*_i \label{gen}
\end{equation}
where $\psi_i$ is the wave function in the scattering region due to
the incoming state $|W_{\alpha,i}>$. From this definition, we can
prove the relationship $(e^2/h) ~ t_{ij} = \int {\bf J}_{ij} \cdot
d{\bf s}$, as follows. It is not difficult to show that the
following ${\bf J}_{i}$ satisfies Eq.(\ref{eq1}): ${\bf
J}_{i}=(ie\hbar/2m) ({\bar \psi}_i \nabla {\bar \psi}_i^* - {\bar
\psi}^*_i \nabla {\bar \psi}_i)-(e^2/m){\bf A} |{\bar \psi}_i|^2$
where ${\bar \psi}_i = \sum_j \psi_j U_{ji} $. Using this ${\bf
J}_{i}$, Eq.(\ref{eq1}) becomes
\begin{equation}
\frac{e^2}{h} ~ t_{i} = \int \sum_{jk} U_{ji} U^*_{ki} {\bf J}_{kj}
\cdot d{\bf s} \ . \label{eq2}
\end{equation}
Denoting ${\cal J}_G$ the generalized current density matrix with
matrix elements ${\bf J}_{ij}$, Eq.(\ref{eq2}) is equivalent to
\begin{equation}
\frac{e^2}{h} ~ {\bar {\cal T}} = U^\dagger \int  {\cal  J}_G \cdot
d{\bf s} U ~~ {\rm or} ~~ \frac{e^2}{h} ~ {\cal T} =  \int {\cal
J}_G \cdot d{\bf s} \ . \label{gen1}
\end{equation}
We have also confirmed Eq.(\ref{gen1}) numerically using specific
examples including that shown in the inset of Fig.\ref{fig1}a.
Therefore, to obtain eigen-current density matrix, we first
diagonalize the transmission matrix ${\cal T}$ to find the unitary
matrix $U$; we then calculate the generalized current density ${\cal
J}_G$ according to Eq.(\ref{gen}). The eigen-current density matrix
is finally obtained by ${\cal J}_{\rm eigen} = U^\dagger {\cal J}_G
U$ and plotted for visualization.

Can eigen-channels be measured experimentally? To answer this
question, as an example let's consider a two-probe device having two
eigen-channels or two edge states in the presence of magnetic field.
Assume one can perform two experiments: (i) measurement of
conductance and (ii) measurement of shot noise. Clearly, conductance
is given by $G=\frac{e^2}{h} (t_1+t_2)$. The shot noise is given
by\cite{buttiker1} $S=\frac{e^2}{h} [t_1(1-t_1)+t_2(1-t_2)]$. From
these $t_1$ and $t_2$ can be determined. In the case of three
eigen-channels, one needs to experimentally measure an additional
quantity\cite{reulet}, for instance the third cumulant of current
$Y=<\Delta {\hat I(t_1)} \Delta{\hat I(t_2)} \Delta{\hat I(t_3)}
>$. In the linear regime, $Y=(e^2/h)\sum_i t_i (1-t_i)
(1-2t_i)$\cite{beenakker}. Hence by measuring $G$, $S$, and $Y$, one
can determine $t_{1,2,3}$. Therefore, the transmission
eigen-channels are physical quantities measurable experimentally.

Having prepared analytical tools, we now present numerical
calculations on how the edge states are destroyed by increasing
degrees of disorder. In the tight-binding representation, the
Hamiltonian of 2D honeycomb lattice of graphene can be written as:
\begin{eqnarray}
H&=&\sum_{i} {\epsilon_{i} c^\dagger_{i} c_{i}} -t
\sum_{<ij>}{e^{i{2\pi} {\phi}_{ij}}}c^\dagger_{i}c_{j}
\end{eqnarray}
where $c^\dag_{i}$ ($c_{i}$) is the creation (annihilation) operator
for an electron on site ${i}$. The first term in $H$ is the on-site
single particle energy where diagonal disorder is introduced by
drawing $\epsilon_{i}$ randomly from a uniform distribution in the
interval $[-W/2,W/2]$ where $W$ measures the disorder strength. The
second term in $H$ is due to nearest neighbor hopping that includes
the effect of a magnetic field. Here the phase ${\phi_{ij}=\int{\bf
A}{\cdot}d{l}/\phi_{0}}$ and $\phi_{0}=h/e$ is the flux quanta. We
fix gauge so that $ {\bf A}=(By,0,0)$; and current flows in the
x-direction. Transmission coefficient is given by $T={\rm Tr}[{\cal
T}]$ where the transmission matrix ${\cal T}$ is obtained from
${\cal T}=\Gamma_R G^{r}\Gamma_L G^{a}$ with $G^{r,a}$ being the
retarded and advanced Green functions of the disordered scattering
region. Quantities $\Gamma_\alpha$ ($\alpha=L,R$) are the line width
functions obtained by calculating self-energies $\Sigma^r$ due to
the semi-infinite leads\cite{lopez84}. The numerical data are mainly
obtained from systems with ${32\times}56$ sites. In the
calculations, energy and disorder strength are measured in unit of
coupling strength $t$.

\begin{figure}
\includegraphics[width=8cm,height=6cm,angle=0]{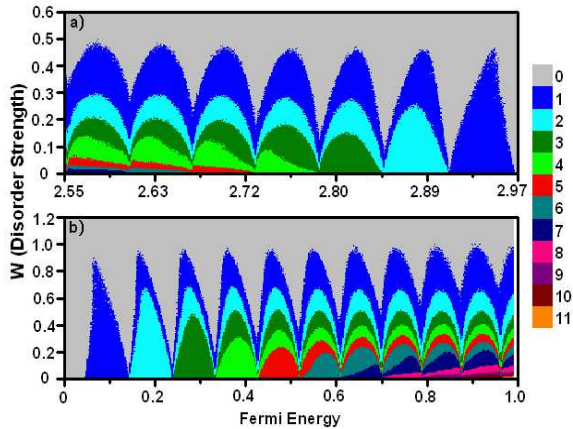}
\caption{(color online) Phase diagram of edge states in $(E,W)$
plane for: (a) hole-like charge carriers in graphene; (b) electrons
in a square lattice. The main difference between the two phase
diagrams is due to opposite charges of hole and electron.}
\label{fig2}
\end{figure}

Numerically, an edge state is identified if transmission coefficient
of an eigen-channel is $T \ge 0.999$; if $T \le 0.001$, the
eigen-channel is said to be closed. In addition, an edge state is
said to be ``destroyed" by disorder and becomes a regular
eigen-channel if its transmission $T$ drops to below $0.999$.
Fig.\ref{fig1}a plots the phase diagram of edge states of graphene
in the $(E,W)$ plane with $\phi=0.0173$ and energy range $[0,0.85]$
where band dispersion is linear (Dirac electrons). A mesh of $600
\times 480$ points are scanned in $(E,W)$ plane and up to 200
disorder configurations are averaged at each point. Several
observations are in order. First, the edge states are destroyed one
by one as $W$ is increased. For instance, at $E=0.5$ the $\nu=3$
edge state is destroyed when $W=0.5$. Very importantly, we emphasize
that at this disorder, there are still three transmission
eigen-channels although only two are edge states and the third being
a regular eigen-channel having $T<0.999$. In other words, the third
eigen-channel is still there to participate transport although it is
no longer an edge state. Increasing disorder to $W=0.7$, the $\nu=2$
edge state is destroyed; finally when $W=1$, all three edge states
are destroyed. We note that edge states would be destroyed all at
once if we had used the usual transmission coefficient for each
channel $t_{ii}$ to characterize the edge states. Second, upon
further increasing $W$, an insulating state is reached where all
eigen-channels are closed. The order of channel closing is also one
by one, in the same order as how edge states are destroyed. This is
shown in Fig.\ref{fig1}c. For instance, at $E=0.5$ there are three
eigen-channels to start with, and at large disorder $W=3.1$, one of
them is closed leaving only two regular eigen-channels. Third, the
edge states are easily destroyed at the subband edges while at the
subband center they are most robust against disorder. This is
because the energy of Landau levels is located at the subband edge.
In the presence of disorder, the Landau level is broadened with a
finite width\cite{Helperin82}. Hence the edge state that is close to
one Landau level can easily relax toward it. Forth, it is more
difficult to destroy an edge state at smaller energies. For Dirac
electrons, the density of states is proportional to $\sqrt{E}$ so
that the level spacing of lower Landau levels is larger than the
upper ones. For electrons with smaller energy it is farther away
from nearby Landau level than electrons with larger energy. Hence a
larger disorder is needed to relax the electrons to the nearby
Landau level. Finally, Fig.\ref{fig1}b plots a ``phase diagram" of
edge states in the $(B,W)$ plane for Dirac electrons. Once again,
the edge states are destroyed one by one, similar to the phase
diagram in the $(E,W)$ plane.

Fig.\ref{fig2}a depicts the ``phase diagram" of edge states of
graphene in $(E,W)$ plane for higher energies in the range
$E=[2.545,2.97]$, where hole-like behavior occurs and band
dispersion is non-linear (non-Dirac electrons). Again, edge states
are destroyed one by one. While the ``phase diagram" topology is
similar to that for Dirac electrons (Fig.\ref{fig1}), here the band
dispersion is quadratic with equal energy spacing between the Landau
levels. Due to this reason, the values of $W$ that are needed to
destroy the last edge state at different subband centers are almost
the same. We have also calculated the ``phase diagram" of edge
states of a square lattice in the $(E,W)$ plane using the same
numerical method, results are shown in Fig.\ref{fig2}b which are
rather similar Fig.\ref{fig2}a. In particular, it is more difficult
to destroy edge states at low filling factor $\nu$, consistent with
the result of Ref.\onlinecite{sheng}.

\begin{figure}
\includegraphics[width=9cm,height=6cm,angle=0]{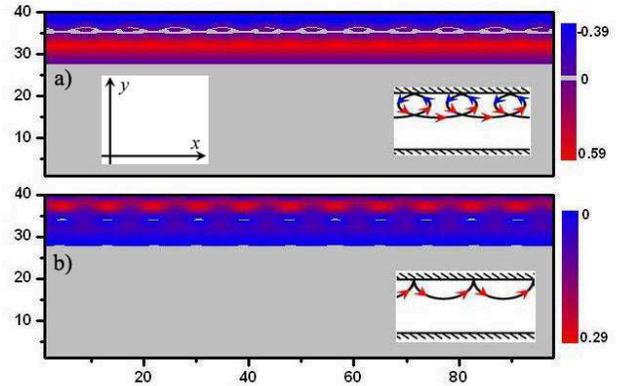}
\caption{(color online) Current density of two edge states flowing
from left to right along x-direction, for a square lattice of
$100\times 40$ sites in the absence of disorder. Here $E=0.4$ and
$\phi=0.052$. Note that the color scale is different for the two
panels. Insets: classical skipping orbits.} \label{fig3}
\end{figure}

\begin{figure}
\includegraphics[width=9cm,height=6cm,angle=0]{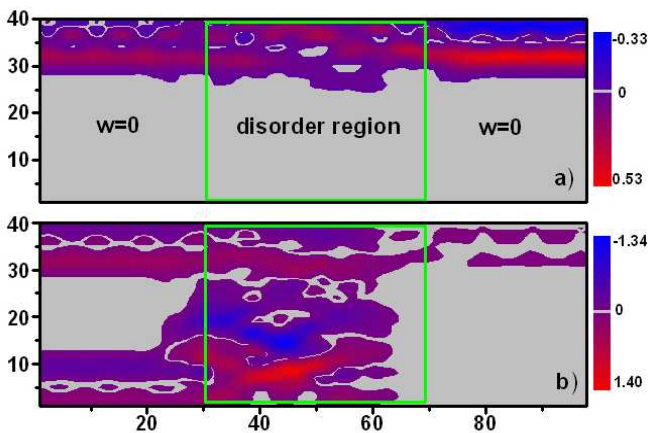}
\caption{(color online) Current density along x-direction for the
square lattice model as that of Fig.\ref{fig3}, with disorder
strength $W=1$, energy $E=0.4$ and magnetic field $\phi=0.052$. (a)
For an edge state that has survived disorder; (b) edge state that
has been destroyed by disorder.} \label{fig4}
\end{figure}

Next, we examine the nature of those edge states that have survived
disorder by calculating current density from Eqs.(\ref{gen}) and
(\ref{gen1}) and plotting it along the propagating direction
(x-direction). Fig.\ref{fig3} shows the current density of two edge
states in the absence of disorder for the square lattice model. Edge
states are clearly seen. Since the two transmission eigen-channels
have different longitudinal energies or effective velocities along
the propagating direction, it gives two different transmission
patterns that correspond to two different skipping orbits of
classical trajectory\cite{beenakker1}. In Fig.\ref{fig3}a, current
flows in the negative direction (blue region) near the sample
boundary and in the positive direction (red region) away from it.
There is a region between these opposite flows where the current
density is very small. The classical trajectory of an electron under
Lorentz force is depicted in the inset, showing a nearly completed
circular motion before colliding with the sample boundary. There is
an one-to-one correspondence between the classical and quantum
motion: near the sample boundary the flow is from right to left,
while it flows opposite away from the boundary. Similar one-to-one
correspondence is also seen in Fig.\ref{fig3}b. For the same square
lattice model, Fig.\ref{fig4} plots the current density of two
eigen-channels for a particular disorder configuration $W=1$ where
the eigen transmission coefficients are $T_1=0.9999$ and
$T_2=0.3385$, respectively. In the numerical calculation, we have
confirmed that the integral of ${\bf J}_i$ over any cross-section
area along the propagating direction gives the same value that is
equal to $(e^2/h) t_i$. From Fig.\ref{fig4}a, it is obvious that
$T_1=0.9999$ is an edge state that survived this degree of disorder.
Compare to Fig.\ref{fig3}, the pattern of current density with
disorder scattering is clearly different. For the eigen-channel with
$T_2=0.3385$, it is clearly a non-edge state (Fig.\ref{fig3}b):
there is a circulating patten with large current density in the
middle of the scattering region, caused by the disorder scattering.
Finally, we have also calculated current density for edge states in
disordered graphene, and similar behaviors are observed as that of
Fig.\ref{fig4}.

In summary, we have investigated the nature of edge states in
disordered mesoscopic samples in the IQHE regime. Our results show
that edge states are destroyed one by one as disorder strength is
increased. In the insulating regime, all transmission eigen-channels
are closed and the closing is also one by one in the same order as
the edge states were destroyed. For graphene and the square lattice
model, the ``phase diagrams" have similar topology but with some
differences due to band dispersions. We have introduced a quantity
which is the generalized current density, using it the current
density of each eigen-channel can be calculated in the presence of
disorder, giving us a vivid physical picture on how edge states are
destroyed. Since transmission coefficients of individual
eigen-channels for mesoscopic samples can be determined
experimentally in the IQHE regime --- as we discussed in the paper,
our conclusions on how edge states are destroyed by disorder should
be testable experimentally.

We note in passing that how to generalize our results to the large
sample limit is an interesting problem requiring further
investigation. In particular, based on the picture of
Khmelnitskii\cite{khmel} and Laughlin\cite{laughlin}, a global phase
diagram of quantum Hall effect in the $(B,W)$ plane was proposed by
Kivelson\emph{et.al}\cite{kivelson} for large samples which
attracted considerable attention both theoretically\cite{ref1,sheng}
and experimentally\cite{exp1}. According to it, an integer quantum
Hall state with a fixed filling factor $\nu$ will \emph{float up} in
energy as the disorder strength increases\cite{khmel,laughlin}. In
the context of \emph{mesoscopic} sample, this idea would mean the
following. Consider Fermi level $E=E_f$, at the mesoscopic sample
boundary there are, say, $\nu$ edge states whose eigen-values cut
this energy $E_f$. The ``float up" idea means that when disorder is
increased, the energies of edge states increase to higher values,
\emph{i.e.} they float up. Hence, at large enough disorder there
will only be $\nu-1$ edge states cutting $E_f$. This way, the system
undergoes a series of transitions between $\nu$ to $\nu-1$ states
\emph{etc.}. Our results presented above, however, indicate that for
mesoscopic samples edge states do not float up by disorder, they are
destroyed to become regular transmission eigen-channels which
participate transport.

\section*{Acknowledgments}
We thank Dr Y.X. Xing for helpful discussions on the classical
analogy of edge states. H.G wishes to thank Prof. X.C. Xie and Prof.
Q. Niu for useful discussions on global phase diagram of quantum
Hall effect. J.W. is financially supported by a RGC grant (HKU
704607P) from HKSAR and LuXin Energy Group. Q.F.S is supported by
NSF-China under Grant No. 10525418 and 10734110; H.G by NSERC of
Canada, FQRNT of Qu\'{e}bec and CIFAR.

\bigskip
\noindent{$^{*)}$ Electronic address: jianwang@hkusua.hku.hk}


\begin{thebibliography}{00}

\bibitem{Helperin82}
B.I. Helperin, Phys. Rev. B {\bf 25}, 2185 (1982).

\bibitem{buttiker88}
M. B\"uttiker, Phys. Rev. B {\bf 38}, 9375 (1988).

\bibitem{prange-book}
See, for example, articles in \emph{The Quantum Hall Effect}, Eds.
R.E. Prange and S.M. Girvin (Springer-Verlag, New York, 1987).

\bibitem{buttiker1}
M. Buttiker, Phys. Rev. B {\bf 46}, 12485 (1992).

\bibitem{Hirai}
S. Komiyama and H. Hirai, Phys. Rev. B {\bf 54}, 2067 (1996).

\bibitem{reulet}
B. Reulet et al, Phys. Rev. Lett. {\bf 91}, 196601 (2003).

\bibitem{beenakker}
M. Kindermann, Yu. V. Nazarov, and C.W.J. Beenakker, Phys. Rev.
Lett. {\bf 90}, 246805 (2003).

\bibitem{lopez84}
M.P. L\'{o}pez-Sancho, J.M. L\'{o}pez-Sancho, and J. Rubio,
J. Phys. \textbf{F} \textbf{14} 1205 (1984).

\bibitem{sheng}
D.N. Sheng and Z.Y. Weng, Phys. Rev. Lett. {\bf 78}, 318 (1997).


\bibitem{beenakker1}
C.W.J. Beenakker and H. van Houten, Solid State Physics {\bf 44}, 1
(1991).

\bibitem{khmel}
D.E. Khmelnitskii, Phys. Lett. A, {\bf 106}, 182 (1984).

\bibitem{laughlin}
R.B. Laughlin, Phys. Rev. Lett. {\bf 52}, 2304 (1984).

\bibitem{kivelson}
S. Kivelson, D.H. Lee, and S.C. Zhang, Phys. Rev. B {\bf 46},
2223(1992).

\bibitem{ref1}
D.Z. Liu, X.C. Xie, and Q. Niu, Phys. Rev. Lett. {\bf 76}, 975
(1996); Th. Koschny \emph{et. al}, Phys. Rev. Lett. {\bf 86}, 3863
(2001); H. Song \emph{et.al}, Phys. Rev. B {\bf 76}, 132202 (2007).

\bibitem{exp1}
J. Glozman \emph{et.al}, Phys. Rev. Lett. {\bf 74}, 594 (1995); T.
Okamoto \emph{et.al}, Phys. Rev. B {\bf 52}, 11109 (1995); S.V.
Kravchenko \emph{et.al}, Phys. Rev. Lett. {\bf 75}, 910 (1995); S.H.
Song \emph{et.al}, Phys. Rev. Lett. {\bf 78}, 2200 (1997); C.H. Lee
\emph{et.al}, Phys. Rev. B {\bf 58}, 10629 (1998); M. Hilke
\emph{et.al}, Phys. Rev. B {\bf 62}, 6940 (2000).


\end{thebibliography}
\end{document}